\documentstyle[12pt]{article}
\begin{document}
\thispagestyle{empty}
\def\cqkern#1#2#3{\copy255 \kern-#1\wd255 \vrule height #2\ht255 depth
   #3\ht255 \kern#1\wd255}
\def\cqchoice#1#2#3#4{\mathchoice%
   {\setbox255\hbox{$\rm\displaystyle #1$}\cqkern{#2}{#3}{#4}}%
   {\setbox255\hbox{$\rm\textstyle #1$}\cqkern{#2}{#3}{#4}}%
   {\setbox255\hbox{$\rm\scriptstyle #1$}\cqkern{#2}{#3}{#4}}%
   {\setbox255\hbox{$\rm\scriptscriptstyle #1$}\cqkern{#2}{#3}{#4}}}
\def\CC{\mathord{\cqchoice{C}{0.65}{0.95}{-0.1}}}
\def\x{\stackrel{\otimes}{,}}
\def\y{\stackrel{\circ}{\scriptstyle\circ}}
\def\proof{\noindent Proof. \hfill \break}
\def\a{\begin{eqnarray}}
\def\b{\end{eqnarray}}
\def\p{{1\over{2\pi i}}}
\def\Q{{\scriptstyle Q}}
\def\P{{\scriptstyle P}}
\renewcommand{\thefootnote}{\fnsymbol{footnote}}

\newpage
\setcounter{page}{0}
\pagestyle{empty}
\centerline{\LARGE Modular Invariance as an Explanation}
\vskip0.3cm
\centerline{\LARGE for the Absence of Monopoles}
\vspace{1truecm} \vskip0.5cm

\centerline{\large F. Toppan}
\vskip.5cm
\centerline{Dipartimento di Fisica}
\centerline{Universit\`{a} di Padova}
\centerline{Via Marzolo 8, I-35131 Padova}
\centerline{\it E-mail: toppan@mvxpd5.pd.infn.it}
\vskip1.5cm
\centerline{\bf Abstract}
\vskip.5cm
It is shown that modular invariance provides a natural explanation
for the absence of monopoles when assumed to be a discrete gauge
symmetry (i.e. the physical states are identified with the orbits
along the modular group).\par
It follows that monopoles can not be seen because
it is always possible to find a suitable gauge-fixing
in which they are not present. This result relies upon
an easy to prove but non-trivial
property of the modular group (a sort of no-monopole theorem for
a diophantine equation).\par
A modular-invariant formulation for the hamiltonian of the
electromagnetism is given. No monopole arises if
independent modular transformations are allowed at each point in
space-time where point-like charges are present. \par
Modular invariance is the quantum analogue of the classical property
that equivalent equations of motion for the electric and magnetic
fields are obtained from any quadratic form hamiltonian
density
$H= a {\vec E}^2 + b{\vec E}\cdot {\vec B} + c {\vec B}^2$, with
$a,b,c$ real, $a,c>0$ and $\Delta= b^2-4ac<0$; it is
therefore quite natural to regard
modular invariance as a gauge symmetry.

\vskip.5cm
\vfill
\rightline{March 1995}
\rightline{DFPD 95-TH-12}

\newpage
\pagestyle{plain}
\renewcommand{\thefootnote}{\arabic{footnote}}
\setcounter{footnote}{0}

\section{Introduction.}

\indent

The modular group $PSL(2,{\bf Z})$ has received recently
a lot of attention in connection with the problem
of the electro-magnetic duality (see e.g.
\cite{oli}-\cite{witt}).\par
In this paper I wish to point out that the modular invariance
provides a natural explanation for both the integral quantization of
the electric charge and the experimental non-observation of the
magnetic monopoles. If modular invariance is regarded as a discrete
gauge symmetry (which means that the physical states in the Hilbert
space are identified with the orbits along the modular group),
then the two properties
above follow immediately. This result is a consequence of an
easy-to-prove but highly non-trivial property of the modular group
(let me call this simple theorem concerning a diophantine equation the
no-monopole theorem; it is demonstrated in the appendix).
Its statement is the following: any two-component vector
${\textstyle{\left( \begin{array}{c} n\\m\end{array}\right)}}$,
with $n,m$ integers ($n$ can be assumed to represent the electric
charge, $m$ the magnetic charge), can be carried with
$PSL(2,{\bf Z})$-modular transformations to a vector of the kind
${\textstyle \left( \begin{array}{c} p\\0\end{array}\right)}$
where $p$ is a uniquely determined integer which labels the
set of inequivalent orbits of the modular group.  \par
{}From this result follows that it is always possible to
find a suitable gauge-fixing in which the monopoles (having
non-vanishing magnetic charge) are not present. The dual nature of
the electric and magnetic field allows defining the electric field
as the one which, in the given gauge-fixing, is associated with the charged
matter; conversely the magnetic field is the one associated
with uncharged matter.\par
The above-cited no-monopole theorem can be regarded just as a funny
property of the modular group with no real physical implications if
the modular invariance is not a symmetry of the dynamics. However
it can be easily provided a modular-invariant formulation for the
hamiltonian of the electro-magnetism. The absence of monopoles in this case
is guaranteed if we allow making independent modular transformations
at each point of the space-time where point-like charges are
present\footnote{In the following discussion Dirac's type monopoles only
are considered.}.\par
In this framework it is more convenient to work with the physical electric and
magnetic fields ${\vec E}, {\vec B}$ instead of the unphysical
gauge-potential $A_\mu$. Manifest Lorentz-covariance is obviously lost
but it is recovered at the level of the equations of motion.\par
The hamiltonian $H$ can be defined to be
\a
H&=& {\textstyle{1\over 2}}\int
d^3{\vec x} (a {\vec E}^2 + b {\vec E}\cdot {\vec B} +c {\vec
B}^2 )
\label{hhh}
\b
(couplings with integral-quantized external sources are also allowed),
where $a,b,c,$ are all assumed to be integers to take into account the
Dirac's integral quantization of the electric and magnetic charges.
Moreover $a$ and $c$ are assumed positive
and the discriminant $\Delta = b^2-4ac<0$ negative
to guarantee the positiveness of the energy ouside the origin. \par
It is possible to compensate the modular transformations of the electric
and magnetic fields ${\vec E}, {\vec B}$ with corresponding modular
transformations acting on $a,b,c$  in such a
way that the hamiltonian $H$ is formally modular-invariant.
In presence of an a priori electrically and/or
magnetically charged matter, the
no-monopole theorem guarantees that it is always
possible to find a gauge-fixing where the magnetic charge is killed.
\par
The fact that modular invariance should be understood as a gauge
symmetry and not as a physical one, follows from this argument:
at the classical level, where no quantization of the charge is
required, the hamiltonian quadratic form (\ref{hhh})
can be assumed depending on (this time) real
coefficients $a,b,c$ satisfying the same properties as above.
The equations of motion derived from any such hamiltonian are equivalent
to the equations of motion for the standard hamiltonian with $a=c=1$,
$b=0$ due to linear canonical transformations redefining the fields
${\vec E}, {\vec B}$. It is therefore natural to regard the modular invariance
of the quantized theory not as a physical invariance acting on
different physical
states, but as a gauge symmetry
connecting different formulations of the same theory
(a brief remark, advocating the possibility of considering
$SL(2,{\bf Z})$ as a discrete gauge group can also be found in
\cite{sen2}).\par
The plan of the paper is the following: in the next section the
properties of the modular group are recalled. It will be commented why
$PSL(2,{\bf Z})$ should be understood as a gauge group and not $SL(2,{\bf
Z})$. The ``no-monopole theorem" will be stated and a modular-invariant
quantum toy-model will be discussed. In the following section
the above results
will be applied to the electromagnetic theory.
The proof of the no-monopole theorem is given in the appendix.

\quad\\
\section{The modular group and a modular-invariant toy-model.}

\indent

The modular group $PSL(2,{\bf Z})$
is defined as the quotient group\\
$PSL(2,{\bf Z}) = SL(2,{\bf Z})/{\bf
Z}_2$, where $SL(2,{\bf Z})$ is the group of $2\times 2$ integer-valued
matrices with unit determinant:
\a
&&{\textstyle{\left( \begin{array}{cc}
a&b\\c&d\end{array}\right)}}\in SL(2,{\bf Z})
\label{mat}
\b
with $a,b,c,d\in {\bf Z}$
and $ad-bc=1$.\\
${\bf Z}_2$ is the subgroup generated by $\pm{\bf 1}_2$.\\
$SL(2,{\bf Z})$ admits two generators which are commonly denoted
as $S,T$:
\a
S &=& {\textstyle{\left( \begin{array}{cc}
0&1\\-1&0\end{array}\right)}};\nonumber\\
T &=&
{\textstyle{\left( \begin{array}{cc}
1&1\\0&1\end{array}\right)}}.
\label{gen}
\b
In mathematics $PSL(2,{\bf Z})$ has a deep geometrical meaning because
it characterizes the inequivalent classes of parabolic (genus $1$)
Riemann surfaces as ${\it H}/PSL(2,{\bf Z})$, where ${\it H}$ is the
Poincar\'{e} upper half plane (see e.g. the references in \cite{num}).\par
In physics, besides the applications to string theory, it is now quite
popular in connection with the Olive-Montonen electric-magnetic duality
realized in four-dimensional supersymmetric theories \cite{witt}.\par
In this section I will discuss some properties of the modular group
and I will use them to define a modular-invariant toy model which can
help clarifying some features shared by more realistic and
phenomenologically
interesting theories.\par
Before doing that, let me just make a short comment on the fact that,
whenever we have a group $G$ of symmetries it must be left to a final
experimental answer the decision whether $G$ acts as a physical
group of symmetries, or if some $G'$ subgroup of it (which
in some cases can even coincide with
$G$ itself) must be considered as a gauge group.\par
To be more precise, let us suppose starting with a
vector space ${\cal V}$ which
carries a representation of $G$ as a left-action, and let us dispose of a
complete set ${\cal O}$ of
{\it observable} operators\footnote{I use this word even if at this stage
no
Hilbert space has been introduced yet; the meaning is simply the
following:
all the eigenvalues for such operators are real, and there exists
a subset of mutually commuting operators
which allows to uniquely determine any vector in ${\cal V}$.}.
Let us
suppose moreover that $G$, due to its adjoint action on ${\cal O}$, acts
as automorphism group for ${\cal O}$. In general many possible Hilbert
space constructions can be performed out of this framework: if we
denote as $|e_n>$ the elements of a basis for ${\cal V}$, a scalar
product can be introduced for ${\cal V}$ and its dual ${\tilde {\cal
V}}$, by defining $<e_m|e_n>=\delta_{mn}$. In this case ${\cal V}$ is a
Hilbert space and $G$ acts as group of physical symmetries,
i.e. it connects different physical states. But we can
also identify the Hilbet space ${\cal H}$ as the coset space ${\cal
V}/G'$ of orbits of ${\cal V}$ along the subgroup $G'$; the scalar
product between elements $|p>\in {\cal H}$ and
$<p'|\in {\tilde{\cal H}}$
being defined now through\\
$
\left\{
\begin{array}{l}
<p'|p> = 1, ~~~if~|p>,~ |p'>~ belong~ to~ the~ same~ G'~
orbit,\\
<p'|p> = 0,~~~ otherwise.
\end{array}
\right.
$~\par
In this case $G'$ is a gauge group. If it is a
normal subgroup of $G$, then $G/G'$ is a factor group which acts
as group of physical
symmetries for ${\cal H}$.\par
It is very well known from the experience we have with the
electrodynamics that physical consequences on ${\cal H}$ can be drawn
from the existence of a gauge symmetry, even when $G'$ coincides with
$G$. In this section I will discuss what happens in such a case, when
$G=G'$ is the modular group.\par
In this case as ${\cal V}$ we can take the space spanned by the vectors
\a
&&\left(\begin{array}{c} n\\m\end{array}\right)~~~with~~n,m~~integers.
\b
The upper element $n$ can be called the electric charge, and the lower
element $m$ the magnetic charge. Under the $S,T$ modular transformations
(\ref{gen})
we get respectively (from now on, for simplicity, I will denote with a
prime the $S$-transformed quantities, with a tilde the $T$-transformed
ones):
\a
\left(\begin{array}{c}n\\m\end{array}\right)&\mapsto&
\left(\begin{array}{c}n'\\m'\end{array}\right) =
\left(\begin{array}{c}m\\-n\end{array}\right)\nonumber\\
\left(\begin{array}{c}n\\m\end{array}\right)&\mapsto&
\left(\begin{array}{c}n\tilde{}\\m\tilde{}\end{array}\right) =
\left(\begin{array}{c}n+m\\m\end{array}\right)
\b
A complete set of commuting operators is given by $N_e$ and $N_m$,
respectively the electric and magnetic
charge number operators, defined through:
\a
N_e\left(\begin{array}{c}n\\m\end{array}\right)&=&
n\left(\begin{array}{c}n\\m\end{array}\right)\nonumber\\
N_m\left(\begin{array}{c}n\\m\end{array}\right)&=&
m\left(\begin{array}{c}n\\m\end{array}\right)
\label{nop}
\b
$N_e,N_m$ transform covariantly under modular transformations, as
it can be easily checked:
\a
\left(\begin{array}{c}N_e\\N_m\end{array}\right)&\mapsto&
\left(\begin{array}{c}N_e'\\N_m'\end{array}\right)=S^{-1}
\left(\begin{array}{c}N_e\\N_m\end{array}\right)=
\left(\begin{array}{c}-N_m\\N_e\end{array}\right)\nonumber\\
\left(\begin{array}{c}N_e\\N_m\end{array}\right)&\mapsto&
\left(\begin{array}{c}{N_e}\tilde{}\\{N_m}\tilde{}\end{array}\right)
= T^{-1}\left(\begin{array}{c}N_e\\N_m\end{array}\right)=
\left(\begin{array}{c}N_e-N_m\\N_m\end{array}\right)
\label{ntra}
\b
A fundamental property of the $PSL(2,{\bf Z})$ modular group has already
been stated in the introduction: it is the content of what can be called
``the no-monopole theorem": each
$\left(\begin{array}{c}n\\m\end{array}\right)$ vector lies in the
$PSL(2,{\bf Z})$-orbit of a vector of the form
$\left(\begin{array}{c}p\\0\end{array}\right)$, where $p$ is an integer
and it is uniquely defined (no other vector
$\left(\begin{array}{c}{\hat p}\\0\end{array}\right)$ with the integer
${\hat p}\neq p$ lies in the same $PSL(2,{\bf Z})$-orbit). Therefore
the whole set of inequivalent $PSL(2,{\bf Z})$-orbits is labelled by
an integer, given by $p$. The proof of this theorem should be found
somewhere in the mathematical literature, however it is so simple that
it is more convenient to show it directly: it is given in the
appendix.\par
Therefore if we regard the set of states connected by $PSL(2,{\bf
Z})$-transformations not as distinct physical states, but as different
gauge-choices of a unique physical state, then we can always
find a suitable gauge-fixing which shows the absence of magnetic
charged matter.\par
Obviously we are free to choose any other gauge fixing (such as
$\left(\begin{array}{c}0\\-p\end{array}\right)$, which is related to the
first one by an $S$-transformation), but in that case we are
also free to redefine our electric and magnetic charge number operators
in such a way that the $N_m$ operator always admits zero-eigenvalues
only. \par
Notice that, even if from a mathematical point of view we can identify
the set of orbits with the integer numbers $p$ (and the $N_m$ operator
is ``needless", so to speak, because of its triviality), on physical
grounds things are not precisely like that:
both the $N_e$ and $N_m$ operators
are (covariantly transforming) physical observables:
indeed monopoles can in principle be actually detected with an
experiment (and up to now $N_m$ has been proven to have only
zero-eigenvalue, apart perhaps the single event found by Cabrera).\par
It is interesting to notice that we could have not considered the full
$SL(2,{\bf Z})$ group as a gauge group. This bigger group admits also
the change in sign, and for that reason
$\left(\begin{array}{c}\pm p\\0\end{array}\right)$ belong to the same
$SL(2,{\bf Z})$-orbit. Therefore in a world in which $SL(2,{\bf Z})$
acts as a gauge group, we have no way to distinguish between positive
and negative values of the electric charge: only the absolute value
of the charge matters. This unphysical property obliges ourselves to
restrict our attention to $PSL(2,{\bf Z})$.\par
There is a priori no reason why $PSL(2,{\bf Z})$ should act as a gauge
group, just as there is no reason why it should act as a physical group
of symmetry. Both possibilities are logically consistent. Obviously it
can happen that only a subgroup of $PSL(2,{\bf Z})$ should be considered
as a gauge group (for instance the discrete subgroup $\Gamma (2)$, see
\cite{num}); in this case restrictions on the allowed spectrum of
monopoles can be drawn.\par
The construction which leads us to the ``no-monopole theorem" allows us
also to introduce a modular-invariant $P$ operator. It is simply
defined through the position:
\a
P \left(\begin{array}{c}n\\m\end{array}\right) &=&
p(n,m) \left(\begin{array}{c}n\\m\end{array}\right)
\b
where $p(n,m)$ is the integer labelling the orbit of
$\left(\begin{array}{c}n\\m\end{array}\right)$ as specified above.
The operator $P$ is therefore modular-invariant by construction.\par
Any function of $P$ which has the right property of being bounded below,
etc., can be considered as a modular-invariant hamiltonian, providing
a modular-invariant dynamics. The simplest, and in some sense most
natural choice for such operators, is to take for hamiltonian
\a
H &=& P^2
\b
When restricted the Hilbert space to the space of $PSL(2,{\bf
Z})$-orbits\footnote{ then $P$ coincides
with $N_e$ in the gauge-fixing specified above.}
we get that, apart the vacuum state, any other energy
state is doubly degenerated, corresponding to the charge-invariance
($P\mapsto -P$) of such a hamiltonian. The spectrum of the theory
is given by the set of squared integers $p^2$ (and coincides, apart
the degeneracy, with the spectrum of the infinite-potential double
well).\par
In this section I have provided the logical ground for the claim
that modular invariance can ``kill" the monopoles. At first
sight it would seem unjustified
to identify the upper and
lower number operators acting on the vectors in ${\cal V}$ with,
respectively, the
electric and magnetic charge number operators. In the next
section I will show that the hamiltonian of the electrodynamics
can be naturally formulated in a modular-invariant framework
and such an identification will no longer appear so arbitrary.
\par

\section{Modular invariance of the electromagnetic hamiltonian.}

\indent

In this section I will furnish a dynamical content to the previous
discussion by showing that the hamiltonian of the electromagnetism
naturally carries a modular invariance when the integral quantization
of the
electric and magnetic charges are taken into account\footnote{Due to
Dirac's result, a quantum theory necessarily involves integral-quantized
electric and magnetic charges (multiples of ${\it e}, {\it b}$
respectively). Conversely an integral charge-quantized theory need not
be a quantum one. In any case I will deserve the name classical for
theories not admitting charge-quantization.}.\par
Before introducing the quantum case, I will discuss the classical
one.\par
Let me at first mention that I will call the Maxwell equations put
in standard form the set of Maxwell equations for the
electric and magnetic fields ${\vec E}, {\vec B}$ with the
normalizations given as follows:
\a
{\dot {\vec E}} &=& {\vec \nabla}\times {\vec B} + {\vec J}\nonumber\\
{\dot {\vec B}} &=& -{\vec \nabla}\times {\vec E} + {\vec K}
\nonumber\\
{\vec{\nabla}}\cdot {\vec E} &=& - J^0\nonumber\\
{\vec \nabla}\cdot {\vec B} &=& -K^0
\label{max}
\b
($J^\mu =J^0, {\vec J}$) and ($K^\mu =K^0,{\vec K}$)
are respectively the electric
and magnetic quadricurrents satisfying the continuity equations
\a
{\dot {J^0}}+{\vec \nabla}\cdot {\vec J} &=&0
\b
(and similarly for ($K^0, {\vec K}$)).\par
The absence of monopoles requires that it is always possible to set
the magnetic quadricurrent equal to zero ($K^0={\vec K} =0$).\par
In the classical case, in absence of matter, the free hamiltonian
which allows deriving the Maxwell equations above with $J^\mu=K^\mu=0$
is given by
\a
H &=& {\textstyle {1\over 2}}\int d^3{\vec x} ({\vec E}^2 +{\vec B}^2)
\label{stham}
\b
where the following Poisson brackets are assumed:
\a
[ E_j ({\vec x},t), B_k ({\vec y}, t) ] &=& i
\epsilon_{jkl}{\textstyle{\partial\over\partial y^l}} \delta ({\vec
x}-{\vec y})
\label{pb}
\b
where $j,k,l = 1,2,3 $ are spatial indices and $\epsilon_{jkl}$ is the
totally antisymmetric tensor
\a
{}~&&(\epsilon_{123}=1).\nonumber
\b
All the other Poisson brackets are assumed vanishing.\par
The group of invariances of the above hamiltonian is rather poor: it is
just given by the parity transformation:
\a
{\vec E}&\mapsto & {\vec E},\nonumber\\
{\vec B}& \mapsto & -{\vec B}
\b
and the $S$-duality
\a
{\vec E} &\mapsto & {\vec B},\nonumber\\
{\vec B} &\mapsto  &-{\vec E}.\nonumber
\b \par
However it should be noticed that the above hamiltonian is not the most
general one satisfying the positivity condition for the energy (outside
the zero-energy solution ${\vec E}={\vec B}=0$).\par
Indeed any hamiltonian
such as
\a
{\hat H}&=& {\textstyle{1\over 2}}\int
d^3{\vec x} (a {\vec E}^2 + b {\vec E}\cdot
{\vec B} + c{\vec B}^2 )
\label{h1}
\b
with $a,b,c$ real numbers, $a,c>0$ and negative discriminant
$\Delta = b^2-4ac<0$, can be used as well. It should also be remarked
however
that $H$ in (\ref{stham}) and ${\hat H}$ in (\ref{h1}) do not provide
different dynamics. It is indeed true that there exists a canonical
transformation for the fields ${\vec E}, {\vec B}$, such as
\a
{\vec E}^{ct} &=& {\sqrt a} {\vec E} + {\textstyle {b\over 2{\sqrt
a}}}{\vec B}\nonumber\\
{\vec B}^{ct} &=& {\textstyle{1\over{\sqrt a}}}{\vec B}
\label{ct}
\b
which transforms the equations of motion for the fields ${\vec E},{\vec
B}$ derived from (\ref{h1}) into the canonical Maxwell equations
(\ref{max}) for the fields ${\vec E}^{ct}, {\vec B}^{ct}$ (in absence
of matter).\par
When taking into account the interaction with matter, and assuming the
integral quantization of the charges, it is natural to restrict
ourselves
to hamiltonian of the form (\ref{h1}), where now the coefficients
$a,b,c$ are assumed integer-valued, and satisfying of course the same
conditions as before:
\a
a,c &>& 0\nonumber\\
\Delta &=& b^2-4ac <0
\label{cond}
\b
In this case ${\hat H}$ easily acquires the structure of a formally
modular-invariant operator; indeed we can compensate the modular
transformations for the vector $
\left(\begin{array}{c} {\vec E}\\{\vec B}\end{array}\right)
$:
\a
\left(\begin{array}{c} {\vec E}\\{\vec B}\end{array}\right)
&\mapsto &
\left(\begin{array}{c} {\vec E}'\\{\vec B}'\end{array}\right)
= S \left(\begin{array}{c} {\vec E}\\{\vec
B}\end{array}\right)\nonumber\\
\left(\begin{array}{c} {\vec E}\\{\vec B}\end{array}\right)&
\mapsto &
\left(\begin{array}{c} {\vec E}~\tilde{}\\{\vec B}~\tilde{}
\end{array}\right)= T\left(\begin{array}{c} {\vec E}\\{\vec
B}\end{array}\right)
\b
(where $S,T$ are given in (\ref{gen}) and ${\vec E}, {\vec B}$ transform
like $n,m$ in the previous section),\\
with corresponding covariant modular transformations on the coefficients
$a,b,c$, in such a way that ${\hat H} $ is left invariant, as it
can be easily checked:
\a
a &\mapsto & a' =c\nonumber\\
b&\mapsto & b' = -b\nonumber\\
c&\mapsto& c' =a
\label{abc1}
\b
and
\a
a&\mapsto & a~\tilde{} = a\nonumber\\
b&\mapsto&  b~\tilde{}=b-2a\nonumber\\
c&\mapsto& c~\tilde{} = a+c-b
\label{abc2}
\b
The above transformations are modular-covariant since
$a,b,c$ transform like
\a
&& a \equiv {N_m}^2;~~~~b\equiv 2N_e N_m;~~~~c\equiv {N_e}^2.
\b
where $N_e,N_m$ are the charge number operators
introduced in (\ref{nop}).\par
Notice that the property (\ref{cond},$1$) is mantained by the above modular
transformations
for $a,b,c$, while the discriminant $\Delta$ in (\ref{cond},$2$) is left
invariant, which implies the consistency of our assumptions.\par
In this framework it is quite evident that the hamiltonian of the
electrodynamics possesses a modular invariance and that such modular
invariance must be understood as a gauge symmetry.\par
Of course we can use our freedom to choose the overall normalization
factor for ${\hat H}$ to set $\Delta = -4$, which corresponds to
the canonical choice (\ref{stham}) for the hamiltonian ($a=c=1$,
$b=0$).\par
Let us discuss now a bit more carefully the equations of motion
derived from the free (\ref{h1}) (or equivalently (\ref{stham})) hamiltonian:
it should be noticed that only the first two Maxwell equations
(those involving the time derivative) are derived
from the Poisson brackets, and the currents are
automatically set equal to zero.\par
The second set of Maxwell equations are not obtained as the
previous ones by making use of the Poisson brackets, rather they are
derived from the first two equations after
taking into account a time-integration.
Therefore the electric and magnetic charge densities $J^0, K^0$
do not appear in the hamiltonian but must be
inserted as boundary conditions. The continuity equations for the
quadricurrents imply that they are distributions of electric and magnetic
charges at rest in the given reference system.
The boundary conditions must be specified according to the principles
we have in mind. Let us imagine having a distribution of point-like
electric and magnetic charges. Taking into account the discussion
of the previous section, and assuming that we can perform independent
modular transformations at each point in the space (for a fixed
time) where
$\left(\begin{array}{c} {n_e}\\{n_m}\end{array}\right)$-electric and
magnetic charges are present (notice that such modular transformations
are indeed invariances of the dynamics, since the charge-distribution
is decoupled from the hamiltonian),
we can therefore conclude that the electric-magnetic charged matter
can be recasted into electric charged matter only. \par
The assumption that independent modular transformations can be performed
is  really a strong assumption, but it is very plausible. It looks
indeed like a sort of principle of general covariance for the
modular transformations or, stated otherwise,
the modular group should be seen as a local discrete gauge group.\par
At this point one can ask what happens when not only charge
distributions at rest, but also currents are taken into account.
The simplest case involves taking an extra-term
\a
m_e {\vec L} \cdot {\vec B} + m_m{\vec L} \cdot{\vec E}&&
\label{int}
\b
in the hamiltonian density (\ref{h1}).
$m_e,m_m$ are integers denoting the integral quantization of the
electric and magnetic charges respectively.\par
The above term (\ref{int}) is modular-invariant provided that
$m_e, m_m$ transform under modular transformations like the $(N_e, N_m)$
charge number operators (\ref{nop}), whose transformations are given by
(\ref{ntra}):
\a
&&(m_e\equiv N_e, m_m\equiv N_m)\nonumber
\b
We can therefore always set the gauge-fixing
where only electric charged matter is present $(m_m=0, m_e=p)$. In such
a gauge the coefficients $a,b,c$ will be generic integers satisfying
the (\ref{cond}) conditions but, as it can be easily checked,
by performing canonical transformations of the ${\vec E}, {\vec B}$
fields, one can derive the standard Maxwell equations (\ref{max}) in
presence of electric current only (${\vec J} = p {\vec \nabla }\times
L$), which is in this case divergenceless.\par
If extra currents are present we can make use of the same principle of
considering the modular group $PSL(2,{\bf Z})$ a discrete
``local" gauge group, to put them in the
same form as above to obtain the vanishing of the magnetic currents.

\quad\\

\section{Appendix: the ``no-monopole" theorem.}

\indent

Let
\a
&&{\hat M }=\left(\begin{array}{cc} a&b\\c&d\end{array}\right)
\nonumber
\b
be a generic matrix in $SL(2,{\bf Z})$.\par
I will fix uniquely the
corresponding element $M$ in $PSL(2,{\bf Z})$ by requiring the following
convention being satisfied: \\
$M$ will be taken the element having $c>0$ if $c\neq 0$; otherwise the one
with $d>0$.\par
Let us prove first that if
$
\left(\begin{array}{c} {p}\\{0}\end{array}\right)$ exists
in the $PSL(2,{\bf Z})$-orbit, then it is
unique; that is easy:
\a
M\left(\begin{array}{c} {p}\\{0
}\end{array}\right)&=&\left(\begin{array}{c} {{\hat p}}\\{0
}\end{array}\right)\nonumber
\b
requires $c=0$ and, for the condition on the
determinant, $ad=1$; since $a,d$ are integers, necessarily\par
\a
a&=&d=1.\nonumber
\b
Let us show now that the equation
\a
M\left(\begin{array}{c} {n}\\{q}\end{array}\right)&=&
\left(\begin{array}{c} {\hat n}\\{0}\end{array}\right)
\nonumber
\b
always admits solutions $M\in PSL(2,{\bf Z})$. \par
Without loss of generality we can take $q>0$.\par
If $n$ is a multiple of $q$: $n=tq$, then many such matrices $M$ can be
found; they are given by
\a
{}~&&\left(
\begin{array}{cc} k & -kt-1\\1 & -t\end{array}
\right)\nonumber
\b
for any integer-valued $k$.\par
If $n$ is not a multiple of $q$, let
\a
n&=&tq + r,\nonumber
\b
with $1\leq r\leq q-1$. \par
Let us now define the
relatively prime positive integers $x,y$ through the position
\a
{x\over y} &=& {r\over q}.\nonumber
\b
We must satisfy the two conditions
\a
ad-bc &=& 1\nonumber\\
cn+dq &=& 0
\nonumber
\b
The second condition tells us that $c$ must be a multiple of $y$:
\a
c &=& ky\nonumber
\b
for some integer $k$.\par
Then
\a
d &= & -k(ty+x)
\nonumber
\b
and inserting these values in the equation for the
determinant we get
\a
 -k[y(at+b) + a x] &=& 1.\nonumber
\b
\par
Necessarily $k=1$ (remember that $c>0$)
and we are reduced to find integer solutions to the equation
\a
my + m' x &=& -1
\nonumber
\b
with $x,y $ relatively prime positive integers. \par
The fact that such
equation always admits solution is of course one of the basic theorem in
number theory (see
however \cite{alg}).

\quad\\

\section{Conclusions.}

\indent

In this paper I have furnished some evidence of the fact that
the modular group $PSL(2,{\bf Z})$ can be considered as a discrete
local gauge group and that, in such a case, some interesting conclusions can
be drawn; in particular the hamiltonian of the electromagnetism it is
shown to be modular-invariant and this naturally lead us, if the above
interpretation proves to be correct, to the absence of monopoles.\par
It would be interesting to understand the above result in a completely
field-theoretical lagrangian framework. The price we have to pay in this
case is the presence of unphysical degrees of freedom, but we could
dispose of a manifestly Lorenz-covariant framework. Indeed duality
invariant actions for the electromagnetism have been discussed in
\cite{{sen},{khp}}.\par
I wish also to point out that it is hardly
conceivable to find a relation between this framework
to explain the absence of monopoles and
the celebrated Yang-Wu \cite{ywu}
result concerning their absence
in trivial fiber-bundles.
In this case we have a discrete symmetry while in the latter
case
the fact that the symmetry group can be continuously deformed plays
a fundamental role. Perhaps, but this is just speculation,
a possible connection can arise if we quantize a $PSL(2,{\bf
R})$-invariant theory.\par
Let me conclude by
noticing that modular-invariant
hamiltonians or actions can be constructed
with a standard procedure. Whenever we find an action
${\cal S}(\tau)$ depending on a complex coupling constant $\tau$
which transforms as
$\tau \mapsto {a\tau +b \over c\tau + d}$, and the action
itself
transforms
covariantly as a modular form of
weight $k$ see\cite{num}, then
we can construct a fully modular-invariant
action by taking
the
norm $||{\cal S}||$, obtained by integrating
${\cal S}{\overline{\cal
S}}$ over the fundamental domain for $\tau$ with the help
of the Petersson metric. In any such case the above considerations can
be applied.

\quad\\

{\large {\bf Acknowledgements.}}

\quad\\

I wish to express my gratitude for helpful discussions and precious
criticism to P. Pasti, D. Sorokin and M. Tonin; I wish also to
acknowledge M. Matone for useful informations on modular stuff.
{}~\\
{}~\\

\end{document}